\documentclass[draft,12pt]{article}
\usepackage{amsmath,amsfonts,palatino,amsthm,epsf}
\newcommand{\beq}{\begin{equation}}
\newcommand{\eeq}{\end{equation}}

\newcommand{\f}{\begin{equation}}
\newcommand{\ff}{\end{equation}}

\begin{document}

\title{Could deformed special relativity naturally arise from the semiclassical limit of quantum gravity?}
\author{Lee Smolin\thanks{Email address:
lsmolin@perimeterinstitute.ca}\\
\\
\\
Perimeter Institute for Theoretical Physics,\\
31 Caroline Street, Waterloo, Ontario N2J 2Y5, Canada, and \\
Department of Physics, University of Waterloo,\\
Waterloo, Ontario N2L 3G1, Canada\\}
\date{\today}
\maketitle

\begin{abstract}
A argument is described for how deformed or doubly special relativity (DSR) may arise in the semiclassical limit of a quantum theory of gravity. We consider a generic quantum theory of gravity coupled to matter, from which we use only the assumption that a Hamiltonian constraint is imposed.  We study circumstances in which $\Lambda, G_N$ and $\hbar$ all may be separately neglected, but there may arise terms in 
the ratio of particle energies to $M_{Pl} = \sqrt{\hbar / G}$ which are small but measurable.  Such cases include probes of an order $M_{Pl}^{-1}$ energy dependent speed of light such as are possible in experiments such as $MAGIC$ and $GLAST$.   We show that in such cases the leading order effect of quantum gravity will, if certain scaling relations are satisfied, be to deform the metric in the effective Hamiltonian of the matter quantum field theory by terms linear in $M_{Pl}^{-1} \times$ energies.  As the Hamiltonian constraint has been imposed there can be no preferred time coordinate or frame of reference, hence this is a modification rather than a breaking of special relativity.

\end{abstract}

\newpage
\tableofcontents

\section{Introduction}

Deformed or doubly special relativity ($DSR$) is a phenomenological proposal for how quantum gravity may leave an imprint on special relativistic quantum field theory in the limit that the familiar effects of gravity are neglected\cite{DSR1,DSR2,DSR-reviews}. It exploits the fact that there is a natural energy and distance scale, the Planck scale, associated with quantum gravity.  The question which DSR addresses is whether, since gravity is a universal phenomena, there might be some universal deformation of quantum field theory parameterized by the Planck mass, $M_{Pl}= \sqrt{\frac{\hbar}{G}}$.  One reason to expect this may be the case is that this constant may survive in the limit in which $\hbar$ and $G$ are both taken to zero. This would leave us with a low energy theory with a symmetry group that preserves not only an observer independent velocity, $c$, but an observer independent energy, $M_{Pl}$. Furthermore, the relativity of inertial observers is kept, there is no breaking of Lorentz invariance and no preferred frame or tensor fields.   In some versions of $DSR$, this invariant energy, $M_{Pl}$ is also the highest attainable for a single quanta (although, obviously, not for a total system.)

The proposal of $DSR$ is especially significant because it is experimentally accessible\cite{DSRphenom}.  DSR theories come in a number of versions and,
generically, predict that the speed of a photon depends on its 
energy, $E$.  This can be made sense of exactly because the constant $M_{Pl}$ lets the dimensionless velocity of a photon depend on the dimensionless ratio $\frac{E}{M_{Pl}}$.  It is expected that the now orbiting GLAST satellite may be able to observe such a variation\cite{GLAST}, and there is even a first claim of a possible observation of this effect by the MAGIC collaboration\cite{MAGIC}.  Given the evidence against Lorentz symmetry breaking, from recent AUGER observations\cite{AUGER}, as well as previously, $DSR$ would be the only explanation were a variation of the speed of photons with energy observed, consistent with the full range of known experimental results.  

There are a number of indications in the literature that $DSR$ does arise in the low energy limit of quantum theories of gravity.  The best understand case is that of quantum gravity in $2+1$ dimensions, where it has been understood for some time that $DSR$ actually characterizes the scattering of classical point particles when coupled to classical general relativity in $2+1$ dimensions\cite{DSR-2+1}.  This arises because of the situation, special to  $2+1$ dimensions, that $\frac{1}{G}$  already has dimensions of mass.  More recently Freidel and Livine have demonstrated that a form of $DSR$ arises as the low energy approximation to  quantum gravity coupled to matter fields\cite{eteralaurent2+1}

While the first argument does not apply in $3+1$ dimensions, it is not so easy to pin the second result on a fact that does not extend from $2+1$ to $3+1$ dimensions.  It is possible to imagine a calculation like that of Freidel-Livine working in $3+1$ dimensions, and this has motivated a certain line of work since.  However, as this rests on technical issues, it is of interest to see if there may be a direct, physical argument, that can lead to the conclusion that $DSR$ is to be expected to arise from a low energy approximation to quantum gravity in $3+1$ dimensions, or explain to us why that is unlikely.  At the very least, one would like an argument that isolates what has to be true of a quantum theory of gravity coupled to matter, for $DSR$ to so arise.

One approach\footnote{Another approach is described in \cite{Carlo-DSR}.}  to such an argument was provided in \cite{falsifiable}.  This provided a criteria for $DSR$ to arise, which concerned the scaling of certain composite operators in the matter theory.  However, this result can also be criticized as being insufficiently general, as it made use of the connection representation of the theory, associated with loop quantum gravity and spin foam models.  The aim of this letter is then to sketch another, apparently independent, approach to the same question, that is more general, and uses only basic features we expect to hold in any quantum theory of gravity.  

Here is an outline of the  basic idea that will be described here. We start from the assertion that to  derive a proper semiclassical limit from quantum gravity we have to start with a nonzero\footnote{Below, assume that $\Lambda >0$, but I know of no reason the argument would not work for negative $\Lambda.$} cosmological
constant $\Lambda$ \cite{positive}. This is motivated by the observation that $L= \Lambda^{-\frac{1}{2}}$ functions as an infrared cutoff for a quantum theory of gravity.  It may then be that quantities of interest in quantum gravity are naturally finite when $\Lambda$ is non-zero, and depend on the dimensionless ratio $\frac{l_{Pl}}{L}$.   To define the phenomenological consequences of quantum gravity in regimes where $\Lambda$ may be neglected we then have to carefully study the limit 
 $\Lambda \rightarrow 0$. 

Some evidence that this limit is involved in $DSR$ was given in \cite{contract-DSR}.  This rests on the fact that 
the $\kappa$-Poincare algebra, which is a quantum deformation of Poincare symmetry associated with $DSR$.  can be understood as the 
$\Lambda \rightarrow 0$ limit of the quantum group
$SO_q(3,2)$.  Here  $q$ must be chosen to scale as a certain function of the cosmological constant.  
\f
q = e^{\frac{\pi \imath }{k+2}},   k = \frac{6 \pi }{G\hbar \Lambda}
\ff
However, results from $LQG$ show that this is exactly how $q$ does depend on $\Lambda$.  

In $2+1$ dimensions this leads directly to the deformation of Poincare symmetry found independently by other methods.  In $3+1$ dimensions it leads to the same scaling criteria for $DSR$ to be relevant as was found in \cite{falsifiable}.

A basic question any derivation of DSR from quantum gravity must answer is how the dependence in just the energy gets into the metric or the deformation of geometry.  In other words, how can there arise quantum corrections that neither satisfy nor break Poincare invariance, but instead deform it?

The answer is in three steps:

\begin{itemize}

\item{}{\bf STEP ONE:}  In the quantum gravity theory the different components of the Einstein's equations 
$G_{\mu \nu} -{\Lambda \over 2} g_{\mu \nu} = G T_{\mu \nu}$ are not treated in the same manner.  The 
$G_{00}$ components make up the Hamiltonian constraint and the
$G_{0i}$ components make up the diffeomorphism constraints.
These are then imposed exactly in the quantum theory. In the path integral formulation these are imposed as $\delta-$functionals, so the path integral is only over histories where the constraints are satisfied at all times.  This is not true of the $G_{ij}$ components of the Einstein's equations which are true only in the classical limit. This is why $T_{00}$ can play a preferred role in the semiclassical limit.

\item{}{\bf STEP TWO:} The Hamiltonian constraint can be solved at each time for the conformal factor of the spatial metric, $\phi$, where the metric is decomposed as 
\f
ds^2 =-  Ndt^2  + N_a dt dx^a + ( e^{\sqrt{\frac{\Lambda}{3}}t   +   \phi}  \delta_{ab} + h_{ab} )  dx^a dx^b 
\ff
where $N$ and $N^a$ are the lapse and shift and $h_{ab}$ is a 
traceless symmetric field representing the gravitational degrees of
freedom.

The Hamiltonian constraint, expressed as an equation for $\phi$, tells us that on each history in the path
integral $\phi$ is determined in terms of the other degrees of freedom by an equation of the form 
$\phi = - G\frac{T_{00}}{\nabla^2 + \frac{\Lambda}{3}}$ where
$T_{00}$ is the sum of the energy momentum tensors of the matter and graviton degrees of freedom.  
This tells us that when the hamiltonian constraint is imposed
the effective metric is of the form
\f
ds^2 =-  Ndt^2  + N_a dt dx^a + ( e^{\sqrt{\frac{\Lambda}{3}}t  - G\frac{T_{00}}{\nabla^2 + \frac{\Lambda}{3}}  }  \delta_{ab} + h_{ab} )  dx^a dx^b 
\label{metric}
\ff

\item{}{\bf STEP THREE:} In the general case the last expression means that the effective low energy metric has no symmetries. But let us consider special cases in which the matter fields are in a state where the $T_{00}$ is homogeneous.  In this case the symmetry
algebra is not broken, it is deformed. We get a realization of DSR in the form in which the metric is energy dependent.

\end{itemize}

This is the basic idea.  To see how it is realized we make two more passes through the idea, the first at an informal level, the second in the context of the path integral for quantum gravity. 

\section{Informal argument}

We consider gravity with positive $\Lambda$ coupled to a very light matter field, $\xi$.  
In the semiclassical limit we can write the spacetime metric in the flat slicing of deSitter spacetime as 
\f
ds^2 = - dt^2 +e^{\sqrt{\frac{\Lambda}{3}}t   +   \phi}  \delta_{ab}  dx^a dx^b 
\ff
where $\phi$ is a small perturbation that carries the gravitational potential. Since it is a small perturbation it satisfies the linearization of the Hamiltonian constraint, which 
in the presence of a cosmological constant $\Lambda= \frac{1}{L^2}$ is 
\f
(\nabla^2 + \frac{\Lambda}{3} ) \phi = G \rho
\ff
where $\rho$ is the matter energy density.
In the semiclassical limit of a quantum theory of gravity this
relation becomes
\f
(\nabla^2 + \frac{\Lambda}{3}  ) \phi = G \hat{T}_{00}
\ff
where $\hat{T}_{\mu \nu}$ is the energy-momentum tensor operator
of the matter quantum field theory.  This implies that
$\phi$ also must be represented by an operator in the matter quantum field theory, defined by 
\f
 \hat{\phi} = G \frac{\hat{T}_{00}}{(\nabla^2 + \frac{\Lambda}{3}  )}
\ff

Suppose the matter field is in a state $|\xi> \in {\cal H}^{matter}$.
Then the effective metric is 
\f
\langle \xi | ds^2 |\xi \rangle  = - dt^2 + 
e^{ \sqrt{\frac{\Lambda}{3}}t    + G \langle \xi |  \frac{\hat{T}_{00}}{(\nabla^2 + \frac{\Lambda}{3}  )} |\xi \rangle }  \delta_{ab} dx^a dx^b 
\label{effectiveg}
\ff
Now let us put the matter field into a one particle state $|k >$ defined
by
\f
\hat{H} |k \rangle = \hbar \omega_k |k \rangle
\ff
where $\hat{H}$ is the matter field Hamiltonian.  We then expect that
\f
\hat{T}_{00} |k \rangle = \frac{ \hbar \omega_k}{L^3}   |k \rangle
\ff
since the cosmological constant functions like an infrared cutoff.  
Furthermore, since the state is a plane wave there is no spatial dependence (in the flat slicing of deSitter) so the terms in
$\nabla^{-2} $ vanish.   

We then have an effective metric operator $\hat{g}_{\mu \nu} (\omega )$ given by 
\f
\langle k | ds^2  | k  \rangle = - dt^2 + 
e^{  \sqrt{\frac{\Lambda}{3}}t    +  \frac{3 G \hbar \omega_k }{(L^3  \Lambda )}  }  \delta_{ab}  dx^a dx^b
\label{meff1}
\ff
Now we want to take the limit $\Lambda \rightarrow 0$.  We have
to remember that so far all quanties in the quantum field have been
bare quantities.  Hence $\omega_k$ is the bare, unrenormalized
frequency of the quantum field.  We know that $l_{Pl}$ is the ultraviolet
cutoff and $L= \Lambda^{-\frac{1}{2}}$ is the infrared cutoff.  Hence
we expect to define a renormalized energy or frequency
\f
\omega_k = Z(\frac{L}{l_{Pl}}) \omega_R
\ff
We expect 
\f
 Z(\frac{L}{l_{Pl}}) =  (\frac{L}{l_{Pl}})^n
\ff
To some power $n$ to be determined so that the limit $\Lambda \rightarrow 0$ exists.  We notice that 
\f
\langle k | ds^2  |k  \rangle = - dt^2 + 
e^{  \sqrt{\frac{\Lambda}{3}}t    +  l_{Pl} \hbar \omega_R Z
\frac{l_{Pl} }{L }  }  \delta_{ab}  dx^a dx^b
\ff
We see that for the limit to exist $n \leq 1$.  If we choose the
limiting value $n=1$ we have in the limit $\Lambda \rightarrow 0$.  
\f
\langle k | ds^2  | k \rangle = - dt^2 + 
e^{  l_{Pl} \hbar \omega_R}  \delta_{ab}  dx^a dx^b
\label{meff2}
\ff
This is a form of an energy dependent metric that gives an energy dependent speed of light.   Its symmetry group is an energy dependent formulation of $DSR$ as given in \cite{DSR2} and \cite{rainbow}.  

Why is this a realization of $DSR$ and not broken lorentz invariance?  The reason is that the Hamiltonian constraint has been implemented and that generates the gauge transformations that correspond to arbitrary changes in the slicing of spacetime, or choice of time coordinate.  Thus, there can be no preferred time coordinate or slicing, hence Lorentz invariance will not be broken in the regime we have been describing. 

For $n < 1 $ we simply get instead the Minkowski metric.  Thus we see that the derivation of DSR appears to involve a critical choice of the scaling of the energy in the presence of the gravitational field. 
This was the case also in \cite{falsifiable} and \cite{contract-DSR}. 

\section{Path integral derivation}

We can reproduce the informal argument just given  via a path integral argument.  We consider gravity with positive $\Lambda$ coupled to matter fields
$\xi$ and $\psi$.  We will be interested in effects of excitations of
$\xi$, the other field $\psi$ goes along for the ride indicating test fields.   The 
full form of the path integral is formally
\f
{\cal Z} (J) = \int dg_{\mu \nu } d\xi d\psi e^{\imath S(g, \xi , \psi ) + J \xi )  }
\delta (\mbox{gauge fixing} ) Det_{FP}
\ff
We will proceed by expanding around deSitter spacetime and then
take the limit $\Lambda \rightarrow 0$.  We are only interested in the leading terms that survive this process.  We decompose the metric according to (\ref{metric}).   As we are interested in just first order terms that arise in the expansion around deSitter spacetime for small $\Lambda$ we write the
lapse as 
\f
N= N_0 + n 
\ff
$n$ and the shift $N^a$ are first order fields. 

We convert the path integral to phase space form and then expand around deSitter spacetime, keeping only leading terms.This means that we integrate over first order deviations $n, N^a$ and $h_{ab}$
as well as the momentum $k^{ab}$.  We also introduce the
momentum $\pi $ and $P$ for $\xi$ and $\psi$.  We have
\begin{eqnarray}
{\cal Z} (J) & = & \int dn \ dN^a \  d\phi \ dh_{ab} \ d\xi \ d\psi \ dk^{ab} d\pi dP
\delta (\mbox{gauge fixing} ) Det_{FP}
\nonumber \\
&&\times 
e^{\imath (k^{ab}\dot{h}_{ab} + \pi \dot{\xi} + P \dot{\psi} -
(N_0+n)  (C_{grav} + T_{00}^\xi + T_{00}^\psi  ) -N^a {\cal D}_a   + J \xi )}
\label{Zexact}
\end{eqnarray} 
Here we have written the Hamiltonian constraint
as
\f
{\cal C} = C_{grav} + T_{00}^\xi + T_{00}^\psi =0
\ff
to show the separation of terms for pure gravity and the two
matter fields.  

To linear order we have
\f
C_{grav} =( \nabla^2 + \frac{\Lambda}{3} ) \phi 
\label{Clinear}
\ff
It is convenient to separate $k^{ab}$ into its trace and tracefree
parts $k, k^{ab}_T$ and to impose the gauge
fixing terms
\f
k=0 , \nabla^a h^{ab} =0
\ff
To this order we also have ${\cal D}_a = \nabla^a k_{ab}$
where indices are raised and lowered by the background deSitter metric. 

We then integrate over the shift $N^a$ and perturbation of the 
lapse $n$ to find the path integral in the form
\begin{eqnarray}
{\cal Z} (J) & = & \int   d\phi \ dh_{ab}^{TT} \ d\xi \ d\psi \ dk^{ab}_T d\pi dP
\nonumber \\
&& \delta \left ( ( \nabla^2 + \frac{\Lambda}{3} ) \phi +T_{00}^\xi + T_{00}^\psi \right )  
\delta (\nabla^a k_{ab}) \delta (k) \delta (\nabla^a h_{ab})
Det_{FP}
\nonumber \\
&+& 
e^{\imath (k^{ab}\dot{h}_{ab} + \pi \dot{\xi} + P \dot{\psi} -
N_0  (C_{grav} + T_{00}^\xi + T_{00}^\psi  )  + J \xi )}
\label{Zlinear}
\end{eqnarray} 

We now integrate over the $\phi$ to solve the linearized 
Hamiltonian constraint.  This gives 
\begin{eqnarray}
{\cal Z} (J) & = & \int  \ dh_{ab}^{TT} \ d\xi \ d\psi \ dk^{ab}_T d\pi dP \ \ 
\delta (\nabla^a k_{ab}) \delta (k) \delta (\nabla^a h_{ab})
Det_{FP}
\nonumber \\
&+& 
e^{\imath (k^{ab}\dot{h}_{ab} + \pi \dot{\xi} + P \dot{\psi} -
N_0  (C_{grav} + T_{00}^\xi (g) + T_{00}^\psi (g)  )  + J \xi )}
|_{\phi =- \frac{T_{00}^\xi  +T_{00}^\psi}{\nabla^2 + \frac{\Lambda}{3} } }
\label{Zlinear2}
\end{eqnarray} 
where the last notation indicates that everywhere the metric
occurs in the above expressions we substitute the conformal
factor $\phi$ by $- \frac{T_{00}^\xi (g^0) +T_{00}^\psi (g^0)}{\nabla^2_0 + \frac{\Lambda}{3} }$.  As we are working only to linear order when we do this we use the background metric with
$\phi =0$ in the expressions for the energy momentum tensor
and $\nabla$.  This is indicated by $g^0$.  

We are not interested in the couplings to gravitons, and their interactions so we ignore and drop all terms in $h_{ab}$ and
$k^{ab}_T$.  We are expanding around a background 
deSitter space which by itself solves $C_{grav}=0$.  In the
gauge we are using $N_0=1$.  
So we are left with simply
\begin{eqnarray}
{\cal Z} (J) & = & \int   d\xi \ d\psi  d\pi dP \ \ 
\nonumber \\
&& 
e^{\imath ( \pi \dot{\xi} + P \dot{\psi} -
 ( T_{00}^\xi + T_{00}^\psi  )  + J \xi )}
|_{\phi =- G \frac{T_{00}^\xi +T_{00}^\psi}{\nabla^2_0 + \frac{\Lambda}{3} } }
\end{eqnarray} 
We can now integrate out the momentum of the matter fields.
We can assume that the matter fields have the standard forms
wheras $T_{00}$ is the same as $H$ with momentum substituted
for time derivatives.  We then have
\begin{eqnarray}
{\cal Z} (J) & = & \int   d\xi \ d\psi  \
e^{\imath  S[\xi, \psi,  g_0 , \phi ]}
|_{\phi =- G \frac{T_{00}^{0 \xi} +T_{00}^{0 \psi}}{\nabla^2_0 + \frac{\Lambda}{3} } }
\end{eqnarray} 

That is, we recover the matter quantum field theory, but on an energy dependent metric defined by
\f
ds^2 = - dt^2 +e^{\sqrt{\frac{\Lambda}{3}}t   - \frac{T_{00}^{0 \xi} +T_{00}^{0 \psi}}{\nabla^2_0 + \frac{\Lambda}{3} }}  \delta_{ab}  dx^a dx^b 
\ff

From this point the argument follows the logic of the section above.  We see that even when $\hbar$ and $G$ can be seperately neglected there is  a residue from quantum gravity which can effect the matter quantum field theory.  This is given by the modified, energy dependent metric above.  Its effect on the matter quantum field theory can then be represented by an effective Hamiltonian in the Fock space of the matter quantum field theory, $\hat{H}_0$.  This is given by the formula
\f
\hat{H}_0(g^0) \rightarrow \hat{H}_{eff} = \hat{H}_0(\hat{g})
\ff
Here, $\hat{g}$ is an operator in the matter Fock space representing an effective energy dependent
metric, which is given by (\ref{effectiveg}).  In the case that we act on a  matter  plane wave state , this gives  (\ref{meff1}), 
\f
\hat{ds^2}  |\psi \rangle  =\left (  - dt^2 + 
e^{  \sqrt{\frac{\Lambda}{3}}t    +  \frac{3 G   \hat{H}_0     }{(L^3  \Lambda )}  }  \delta_{ab}  dx^a dx^b
\right )  |\psi \rangle
\label{meff3}
\ff
We now take  the limit $\Lambda \rightarrow 0$ on a basis of plane wave states.  
When the limit is taken as described above we reach the form (for the critical case $n=1$)
\f
ds^2  | \psi  \rangle = \left ( - dt^2 + 
e^{  l_{Pl}  \hat{H}_0  }  \delta_{ab}  dx^a dx^b \right )  | \psi  \rangle
\label{meff4}
\ff

That is, the effective spatial part of the metric is now an operator, given by
\f
\hat{q}_{ab}= \delta_{ab}e^{  l_{Pl}\hat{H}_0 }
\ff
This gives a universal modification of the matter perturbation theory.  For example,  we note that one consequence is that the scalar field propagator is
modified so that
\f 
\frac{i}{E^2 - p^2 - m^2}  \rightarrow \frac{i}{E^2 - p^2e^{-l_{Pl}E} - m^2}
\ff

Thus, the residue left by quantum gravity is to deform the metric and make it energy dependent.   
We note that this must be a deformation rather than a breaking of the Poincare algebra as it arose from the implementation of the Hamiltonian constraint, hence a preferred frame cannot have been introduced. 

We also note that the derivation assumes that there are not other quantum gravity effects that must be taken into account in the effective Hamiltonian.  This requires that the total energy be small compared to the Planck energy.  As a result, the derivation, as it stands, does not address the issue of what happens for states with energy large in Plank units.

\subsection*{Perturbation theory in the presence of the deformed metric}

We can easily see that the gravitons can play the role of the matter field in deforming the background metric.  Let us go back to
(\ref{Zexact}) but include the second order term in gravitons in
the expansion of the Hamiltonian constraint.  In place
of (\ref{Clinear}) we have
\f
C_{grav} =( \nabla^2 + \frac{\Lambda}{3} ) \phi + G\tau_{00}^h  
\label{Csecond}
\ff
where $\tau_{00}^h$ is the quadratic term in $h$ of the energy-momentum pseudo-tensor of the gravitational field. This functions just like the energy momentum tensor of the graviton field.   
The path integral then treats the gravitons like a matter field,
The partition function (\ref{Zlinear}) now has the graviton
term added
\begin{eqnarray}
{\cal Z} (J) & = & \int   d\phi \ dh_{ab}^{TT} \ d\xi \ d\psi \ dk^{ab}_T d\pi dP
\nonumber \\
&& \delta \left ( ( \nabla^2 + \frac{\Lambda}{3} ) \phi +G T_{00}^\xi + G T_{00}^\psi+ G \tau^h_{00}\right )  
\delta (\nabla^a k_{ab}) \delta (k) \delta (\nabla^a h_{ab})
Det_{FP}
\nonumber \\
&+& 
e^{\imath (k^{ab}\dot{h}_{ab} + \pi \dot{\xi} + P \dot{\psi} -
N_0  (C_{grav} + GT_{00}^\xi + G T_{00}^\psi  + G \tau^h_{00})  + J \xi )}
\label{Zquadratic}
\end{eqnarray} 
Integrating over $\phi$ we have now
\begin{eqnarray}
{\cal Z} (J) & = & \int  \ dh_{ab}^{TT} \ d\xi \ d\psi \ dk^{ab}_T d\pi dP \ \ 
\delta (\nabla^a k_{ab}) \delta (k) \delta (\nabla^a h_{ab})
Det_{FP}
\nonumber \\
&+& 
e^{\imath (k^{ab}\dot{h}_{ab} + \pi \dot{\xi} + P \dot{\psi} -
N_0  (C_{grav} + G T_{00}^\xi (g) + G T_{00}^\psi (g) + G \tau^h_{00} )  + J \xi )}
|_{\phi =- G \frac{T_{00}^\xi  +T_{00}^\psi +  \tau^h_{00}}{\nabla^2 + \frac{\Lambda}{3} } }
\label{Zquad2}
\end{eqnarray} 

Thus, we see that the gravitons also propagate on the deformed, energy dependent spacetime geometry.   

\section{Conclusions}

We have presented here the outline of a derivation of $DSR$ from quantum gravity.  It points to a new kind of modification or deformation of quantum field theory in which new non-linearities are introduced by making the metric of the background spacetime dependent on the energy density or  total Hamiltonian of the original, non-gravitational quantum field theory.  

We showed here that this proposal follows from  three assumptions.  First that the limit of quantum gravity in which $\hbar$ and $G$ are both small, is to be taken in such a way that their ratio, $M_{Pl}$ is held fixed. Second, that the quantum gravity theory must be defined with $\Lambda$ fixed, and the limit of $\Lambda \rightarrow 0$ taken carefully and, third, that for leading order effects in these limits, the Hamiltonian constraint of the quantum gravity theory can be imposed in the Hilbert space of a matter quantum field theory, making the gravitational potential an operator in the matter Hilbert space.  An argument for the latter was given in terms of the semicassical limit of the path integral in a quantum gravity theory.   

This argument provides a justification for the view that $DSR$ may be understood as an energy dependent deformation of the geometry of spacetime, in the limit that gravity is turned off\cite{rainbow}.   It also explains what the energy is in that proposal: it is the total Hamiltonian of the matter quantum field theory, without taking gravity into account.  While this is helpful, more work is clearly needed to develop the implications of this hypothesis and work out the details of the modifications of quantum field theory sketched here.  

Important issues remain to be addressed, including the soccer ball problem and the issue of the effect of any non-localities.  At the same time, the sketch of a derivation given here suggestions new approaches to resolving these issues.  The first priority remains the experimental test of the basic prediction of $DSR$, which is an energy dependent speed of light, while maintaining the relativity of inertial frames.  

\section*{Acknowledgements}

I am grateful to Giovanni Amelino-Camelia, Laurent Freidel,  Jerzy Kowalski-Glikman and Chanda Prescod-Weinstein for  conversations related to DSR over the last years.  A first attempt at the above argument arose in collaboration with Carlo Rovelli, his own approach to this problem is presented in \cite{Carlo-DSR}.  I am grateful finally to Sabine Hossenfelder for many helpful comments on a draft of this paper and to Joao Magueijo for collaboration and many conversations about DSR and in particular for his encouragement to finish this paper. Research at
Perimeter Institute is supported in part by the Government
of Canada through NSERC and by the Province of Ontario through MEDT.

\end{document}